\documentclass[aps,reprint,superscriptaddress]{revtex4-1}
\usepackage{mathtools}
\usepackage{amsmath,amssymb}
\usepackage{color}
\usepackage{braket}
\usepackage{graphicx}
\usepackage{epstopdf}
\usepackage{float}
\usepackage{nccmath}
\usepackage[normalem]{ulem} %\sout
\usepackage{textgreek}
\usepackage{colortbl}
\usepackage{multirow}
\usepackage{hhline}

\usepackage{soul}
\usepackage[table]{xcolor}

\begin{document}

\title{Tracking exceptional points above laser threshold}

\author{Kaiwen Ji}
\affiliation{Centre de Nanosciences et de Nanotechnologies, CNRS, Université Paris-Saclay, 10 Boulevard Thomas Gobert, 91120 Palaiseau, France}

\author{Qi Zhong}

\affiliation{Department of Physics, Michigan Technological University, Houghton, Michigan 49931, USA}
%\affiliation{Henes Center for Quantum Phenomena, Michigan Technological University, Houghton, Michigan 49931, USA}

\author{Li Ge}
%\email[]{li.ge@csi.cuny.edu }
\affiliation{Department of Physics and Astronomy, College of Staten Island, CUNY, Staten Island, New York 10314, USA and Graduate Center, CUNY, New York, New York 10016, USA}

\author{Gregoire Beaudoin}
%\email[]{alejandro.giacomotti@c2n.upsaclay.fr }
\affiliation{Centre de Nanosciences et de Nanotechnologies, CNRS, Université Paris-Saclay, 10 Boulevard Thomas Gobert, 91120 Palaiseau, France}

\author{Isabelle Sagnes}
%\email[]{alejandro.giacomotti@c2n.upsaclay.fr }
\affiliation{Centre de Nanosciences et de Nanotechnologies, CNRS, Université Paris-Saclay, 10 Boulevard Thomas Gobert, 91120 Palaiseau, France}

\author{Fabrice Raineri}
%\email[]{alejandro.giacomotti@c2n.upsaclay.fr }
\affiliation{Centre de Nanosciences et de Nanotechnologies, CNRS, Université Paris-Saclay, 10 Boulevard Thomas Gobert, 91120 Palaiseau, France}

\author{Ramy El-Ganainy}
%\email[]{ganainy@mtu.edu}
\affiliation{Department of Physics, Michigan Technological University, Houghton, Michigan 49931, USA}
\affiliation{Henes Center for Quantum Phenomena, Michigan Technological University, Houghton, Michigan 49931, USA}

\author{Alejandro M. Yacomotti}
%\email[]{alejandro.giacomotti@c2n.upsaclay.fr }
\affiliation{Centre de Nanosciences et de Nanotechnologies, CNRS, Université Paris-Saclay, 10 Boulevard Thomas Gobert, 91120 Palaiseau, France}

\begin{abstract}
Recent studies on non-Hermitian optical systems having exceptional points (EPs) have revealed a host of unique characteristics associated with these singularities, including unidirectional invisibility, chiral mode switching and laser self-termination, to mention just a few examples. The vast majority of these works focused either on passive systems or active structures where the EPs were accessed below the lasing threshold, i.e. when the system description is inherently linear. In this work, we experimentally demonstrate that EP singularities in coupled semiconductor nanolasers can be accessed and tracked above the lasing threshold, where they become branch points of a nonlinear dynamical system. Contrary to the common belief that unavoidable cavity detuning will impede the formation of an EP, here we demonstrate that this same detuning is necessary for compensating the carrier-induced frequency shift, hence restoring the nonlinear EP in the lasing regime. Furthermore, unlike linear non-Hermitian systems, we find that the spectral location of EPs above laser threshold varies as a function of total pump power and can therefore be continuously tracked. Our work is a first step towards the realization of lasing EPs in more complex laser geometries, and enabling  the enhancement of photonic local density of states through non-Hermitian symmetries combined with nonlinear interactions in coupled laser arrays.
\end{abstract}

\maketitle

\section*{Introduction}
Exceptional points (EPs) are algebraic branch points associated with multi-valued complex functions. In physics, EPs are associated with the spectra of non-Hermitian systems. Despite the early theoretical studies on EPs \cite{Heiss2012JPA,Rotter2003PRE} and experimental efforts to demonstrate some of their features using microwave setups \cite{Dietz2007PRE,Dietz2011PRL}, it was not until the seminal work on parity-time (PT) symmetric potentials in quantum mechanics  \cite{Bender1998PRL,Bender1999JMP}, and its introduction to optics \cite{Ganainy2007OL,Makris2008PRL,Musslimani2008PRL,Ruter2010NPhys} that the notion of EPs in physics has attracted considerable attention, in large part due its potential applications in optics and photonics. For recent reviews, see \cite{Feng2017NP,Ganainy2018NP,Ozdemir2019NM,Miri2019S,Wiersig2020PR}.  

Among the variety of optical platforms where EPs and their ramifications can be investigated, laser systems are particularly interesting due to the flexibility in engineering their non-Hermiticity (by adding gain and loss at will) and the ability to control their nonlinearities (by choosing the appropriate material and adjusting the pump levels). This unique combination of features, coupled with the well-developed experimental techniques for measuring laser characteristics and applying different feedback schemes to control their operation, have enabled researchers to use various laser setups as a test bed for exploring a number of intriguing physical effects such as wave chaos \cite{Sciamanna2015NP}, Anderson localization of light \cite{Liu2014NN} and symmetry breaking \cite{Hamel2015NP,Ghofraniha2015NC}.

In recent years, several experimental studies have demonstrated how PT symmetry and EPs can be utilized to control the lasing modes in multimode laser arrangements \cite{Miri2012OL,Hodaei2014S,Feng2014S,Peng2016PNAS,Miao2016S}. Subsequent theoretical works have elaborated more on the nonlinear dynamics of these systems \cite{Ge2016SR,Teimourpour2017SR,Hassan2015PRA,Kominis2018APL,Kominis2017PRA}. An interesting feature associated with the presence of EPs in laser systems is that of laser self-termination where applying a spatially inhomogeneous pump to a lasing device can shut down the laser action altogether \cite{Liertzer2012PRL,Brandstetter2014NC}. Conversely, applying an inhomogeneous loss to a non-lasing device can lead to lasing \cite{Peng2014S}. In almost all this aforementioned work, the emphasis was on approaching EPs below the lasing threshold. In fact, it was explicitly demonstrated in  \cite{Ganainy2014PRA} that laser self-termination (or loss induced lasing) can take place only under that condition. 

EPs above laser threshold have recently been investigated theoretically, predicting that the EP laser can be stable for a large enough inversion population relaxation rate \cite{https://doi.org/10.48550/arxiv.2206.12969}. From the experimental point of view, however, even in the more recent works on PT symmetric laser \cite{Hassan2015PRA, Kim2016NC,Gao2017O}, the relation between the lasing characteristics and the relative position of the EP with respect to the lasing threshold was not studied. And while sensing devices based on a laser operating at a third order EP were presented in \cite{ganainy2017Nature}, and the signature of crossing an EP above the lasing threshold was reported in \cite{Brandstetter2014NC}, these systems were only analyzed within the linear coupled mode equations, which cannot capture the inherent nonlinear dynamics around EPs, as a result of complex mode bifurcation and stability. 
In addition, the interplay between the nonlinear frequency shift induced by the amplitude phase coupling in semiconductor lasers and the onset of EPs has received very little, if any, attention. In fact, it was concluded in~\cite{takata2021observing} that this nonlinear frequency shift, together with the unavoidable cavity detuning due to fabrication errors, will impede the formation of EPs in the lasing regime because of the narrow linewidths in play, and thus EPs can only be closely approached below the laser threshold.

In this work, we report on the observation of EPs above laser threshold; we investigate the conditions for their existence and how they impact the lasing characteristics of semiconductor cavities. In particular, we study a photonic molecule laser made of two coupled photonic crystal nanocavities and characterize the emission as the applied differential gain as well as the total power are varied. Our main finding is that the carrier-induced frequency shift breaks the effective PT symmetry of the system and thus removes the EP; however, the EP singularity can be restored by introducing an opposite frequency detuning at the fabrication stage (see Fig.~\ref{Fig_Bifurcation1}(d)). Furthermore, above the lasing threshold, the location of EP in terms of the differential gain depends on  the total applied gain.
This last feature is very different from the usual scenario where EPs are approached below lasing threshold, the system becomes linear and the EP is pinned down by the value of the coupling between the two cavities. As we will show shortly, this distinction between the behavior of the EP below and above the lasing threshold will be instrumental for characterizing the system under investigation. Our work provides more insight into the interplay between EPs and nonlinear interactions in coupled semiconductor nanolaser systems under nonuniform pumping. As such, it serves as a first step towards investigating more complex laser networks under extreme non-Hermitian and nonlinear conditions, such as those involving random lasers with large number of modes \cite{Tureci2008S,Schonhuber2016O,Hisch2013PRL}.

\section*{Results}

\begin{figure*}[!t]
%\centering
\includegraphics[width=\linewidth]{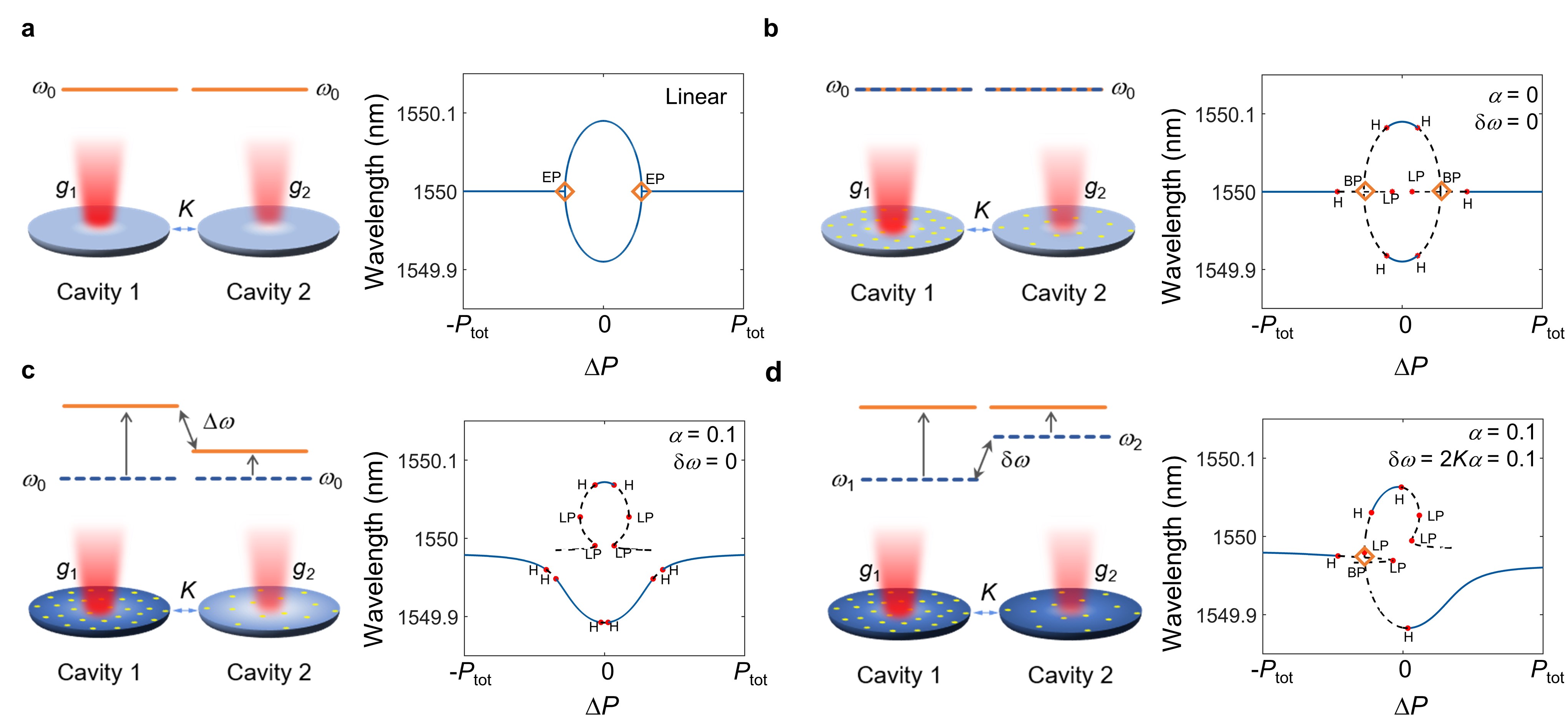}
	\caption{\textbf{Concept of lasing EP.}
	 For reference, a linear PT symmetric coupled-cavity system is presented in (a). The case of coupled identical laser cavities with an ideal gain medium is plotted in (b): here nonlinear saturation effects are taken into account but the population inversion does not induce any refractive index change, therefore there is no pump-induced frequency shift of the resonant modes ($\alpha=0$ in Eqs. \ref{Eq-LaserRateEq}). Note that the EP becomes a nonlinear (pitchfork) bifurcation point (marked by the label BP). (c) A nonzero phase-amplitude coupling in the semiconductor cavity (nonzero $\alpha$-factor) induces an asymmetric blue shift, with a net detuning $\Delta \omega$ between the two cavities and thus breaks their parity symmetry. As a result, it impedes the formation of the EP. The effective PT symmetry ($g_1=-g_2, g_j=-\kappa+(n_j-n_0)\beta\Gamma_{\parallel}/2, j=1,2$), and hence the formation of an EP can be restored if the (active) blue shift is compensating for by using a static red shift (i.e. introduced in the design from the beginning) $\delta \omega$ as shown in (d). The colors of the disks represent the sum of the frequency shift due to the carriers, $\Delta\omega_{1,2}=\alpha(n_{1,2}-n_0)\beta\Gamma_{\parallel}/2$, and the fabrication. Here we take  $P_{tot}=3P_0$, where $P_0$ is the threshold of a single cavity. In the bifurcation diagrams, H represents the Hopf bifurcation and LP is the limit point.}
	\label{Fig_Bifurcation1}
\end{figure*}

\begin{figure}[!t]
%\centering
\includegraphics[width=\linewidth]{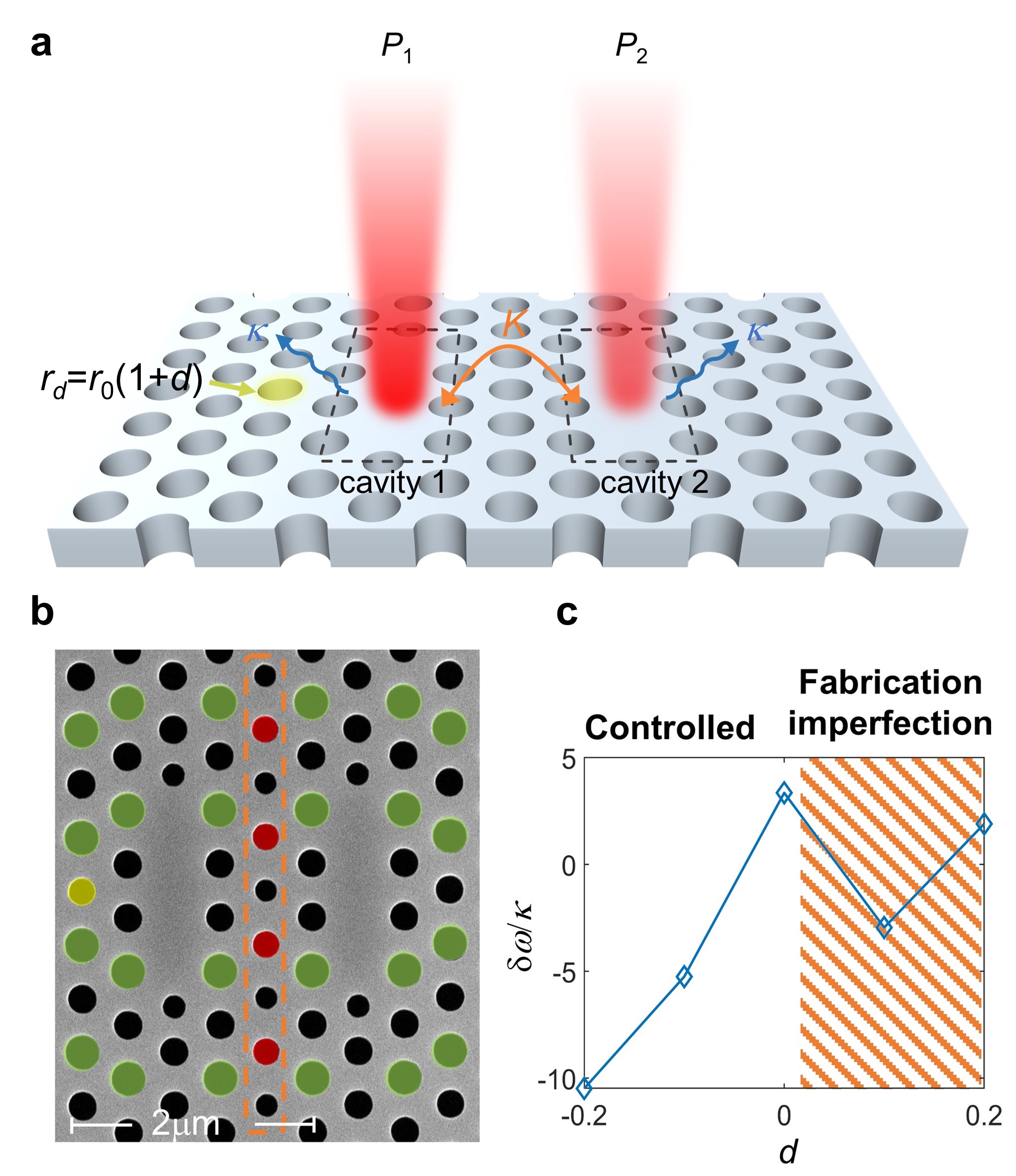}
	\caption{\textbf{Design and fabrication of coupled nanolasers.} \textbf{a} A schematic of the photonic molecule laser under investigation. It consists of two coupled photonic crystal nanocavities. The coupling strength between the two cavities can be controlled by changing the radii of the nanoholes that lie exactly at the center between the two cavities. This changes the resonant frequencies as well but by the same amount in both cavities. In turn, the individual cavity frequencies of the cavities are independently tuned by changing the size of a neighboring hole (highlighted in yellow close to cavity 1). Finally,  two pump beams with different intensities providing pump rates $P_{1,2}$, are used to provide unequal gain for the two nanocavities. \textbf{b} SEM image of the fabricated sample: the material is InP with embedded InGaAsP quantum wells. The lattice constant $a=430$nm, $r_0=0.266a$, $h=-0.25$ and $d=-0.1$. The coupling is controlled by the barrier between two cavities, which is displayed in orange dashed box ($r_b=(1+h)r_0$). The yellow hole is the detuning hole ($r_d=(1+d)r_0$). The brown boxes indicate engineered holes to improve the beaming-quality of the radiated photons ($r_{beaming}=r_0+0.05a$). The overlaps between the beaming holes and the barrier, colored in red, have radii of $r'=r_{beaming}(1+h)$. \textbf{c} Detuning as a function of radii $d$. Fabrication imperfections are dominant for $d>0$, therefore we restrict our studies to the range of $-0.2\leq d\leq 0$, where the cavity-detuning is well-controlled by design.}
	\label{Fig_Schematic}
\end{figure}

\textit{EPs in coupled semiconductor lasers:---}Figure \ref{Fig_Schematic} depicts a schematic of the photonic molecule laser under consideration. It consists of two coupled photonic crystal cavities implemented on a InP-based standing membrane with embedded quantum wells. This platform has been used recently for investigating physical effects such as spontaneous symmetry breaking \cite{Hamel2015NP,PhysRevLett.128.053901}, superthermal light generation \cite{PhysRevX.8.011013}, and mesoscopic limit cycles \cite{Marconi2020PRL}. In principle, the resonant frequency of each cavity and their coupling coefficient can be controlled by carefully engineering the cavity area, the separation between the two cavities and the size of the nano-holes. In our design however, we only tune the resonant frequency of cavity 1 and the coupling coefficient (detailed discussion on the design parameters will be presented later).

The lasing action of the above system can be well-described by the following rate equation model that accounts for both the field and carrier dynamics:

\begin{subequations} \label{Eq-LaserRateEq}
\begin{align}\
\begin{split}
\frac{da_{1,2}}{dt}&=\left[i\omega_{1,2}-\kappa+\frac{1+i\alpha}{2}\beta\Gamma_{\parallel}(n_{1,2}-n_0)\right]a_{1,2}\\
&+iK a_{2,1} + F_{1,2}(t)
\end{split}
\\
	\frac{dn_{1,2}}{dt}&=P_{1,2}-\Gamma_{tot}n_{1,2}-\beta\Gamma_{\parallel}(n_{1,2}-n_0)\left|a_{1,2}\right|^2,
\end{align}
\end{subequations}

The various variables and parameters in Eq (\ref{Eq-LaserRateEq}) are listed in Table I.\\  

\begin{table}[!b] 
\caption{List of the variables and parameters used in  Eq (\ref{Eq-LaserRateEq})} 
\centering
\begin{tabular}{|p{0.6in}<{\centering}|p{2.7in}|} 
  \hline
  \multicolumn{1}{|c|}{\bf Symbol} & \multicolumn{1}{|c|}{\bf{Physical quantity}}\\ 
  \hline
  $a_j$ & \:Field amplitudes in cavity $j$ \\ 
  \hline
  $\omega_j$ & \:Resonant frequency of cavity $j$ \\ 
  \hline
  $K$ & \:Coupling coefficient between the two cavities \\ 
  \hline
  $\kappa$ & \:Cavity loss rate \\ 
  \hline
  $n_0$ & \:Carrier number at transparency \\ 
  \hline
  $n_j$ & \:Carrier number in cavity $j$ \\ 
  \hline
  $\alpha$ & \:Phase-amplitude coupling (also known as linewidth enhancement) factor \\ 
  \hline
  $\beta$ & \:Spontaneous emission coefficient \\ 
  \hline
  $\Gamma_{\parallel}$ & \:Two-level radiative recombination rate \\ 
  \hline
  $\Gamma_{tot}$ & \:Total carrier recombination rate \\ 
  \hline
  $F_{j}(t)$ & \:Langevin noises  \\ 
  \hline
  $P_{j}(t)$ & \:Pump rate in cavity $j$  \\ 
  \hline
\end{tabular}
\end{table}

As indicated above, the two cavities have identical loss coefficients. The pump-imbalance $\Delta P$, defined as $\Delta P=P_1-P_2$, is the non-Hermitian control parameter. Before we present the experimental results, it is instructive to first plot the lasing frequencies obtained from  Eq (\ref{Eq-LaserRateEq}) for different values of the parameter $\alpha$. As a reference, we first plot in Fig.\ref{Fig_Bifurcation1}(a) the linear case, i.e. in the absence of gain and loss saturation nonlinearities (terms proportional to $|a|^2$ in Eq. \ref{Eq-LaserRateEq}b neglected) and carrier-induced frequency shift ($\alpha=0$), which reduces to the standard EP bifurcation. When the gain/loss saturation nonlinearity are included but we still take $\alpha=0$, we observe that the eigenfrequency branching now takes place across a pitchfork bifurcation  as shown in Fig.\ref{Fig_Bifurcation1}(b). The dashed/solid lines indicate unstable/stable lasing modes. On the other hand, when the value of $\alpha$ is finite (in that case $\alpha= 0.1$), the EP disappears (see Fig.\ref{Fig_Bifurcation1}(c)). This can be easily explained by the carrier-induced blueshift that breaks the PT symmetry of the system. If, however, we introduce a linear frequency detuning $\delta \omega\equiv (\omega_1-\omega_2)|_{ext}$ ---the subscript here indicates that this frequency detuning is introduced by external means, for instance in the design parameters--- that compensate for the nonlinear frequency shift, the EP can be restored again as shown in Fig.\ref{Fig_Bifurcation1}(d). We also point out that, in some regions around the bifurcation points, all the modes are unstable. As a matter of fact, time domain integration of Eq (\ref{Eq-LaserRateEq}) shows that the laser output along these unstable branches is oscillatory as opposed to stable steady states. %From a practical perspective, this means that these branches can have a measurable contribution to the lasing action \textcolor{orange}{AG: not clear} \textcolor{red}{RE: I agree, this sentence is not clear}.

As we show in SM, even for realistic values of $\alpha=2-5$, compensating for the nonlinear frequency shift through a linear frequency detuning is still possible. In fact, the carrier-induced blue-shift is given by $\Delta \omega=\alpha \beta\Gamma_{\parallel}(n_1-n_2)/2$ and the linear frequency detuning required to compensate for this value is simply $\delta \omega=-\Delta \omega$. Assuming $n_1>n_2$ without loss of generality, at the onset of the EP, the magnitude of the required linear shift is given by $\delta\omega= -2K\alpha$; note that, in this case, the sign indicates that cavity 1 must be externally red-detuned to compensate for the carrier-induced blue-shift (see SM note \ref{Sec_Compensation} for detailed derivation of the above formulas).
Importantly, from  Eq. (\ref{Eq-LaserRateEq}), one can also show that, above the lasing threshold, the pump imbalance at the EP is a function not only of the gain difference between the cavities $\Delta g_{2,1}$, but it also varies with the total pump $P_{tot}$,  $\Delta P|_{EP}=-\Delta g_{2,1}|_{EP}(P_{tot}-2n_0\Gamma_{tot})/2\kappa$ (see SM note \ref{Sec_DeltaP}). This can be understood intuitively by recalling that the onset of an EP is determined by the gain difference between the two resonators, which is reduced for increasing intracavity intensity. Consequently, the effective gain difference saturates and the pump imbalance needs to be larger to reach the coupling rate at the EP. An interesting outcome of our nonlinear analysis with no counterpart in linear systems is that $\lim_{P_{tot}\rightarrow{\infty}}\Delta P|_{EP}/P_{tot}=K/\kappa$. Thus, in order to observe an EP under arbitrary pump conditions, the design parameters of the coupled nanocavities must satisfy the weak intercavity coupling condition, $K< \kappa$.

\begin{figure*}[!t]
	\centering
	{\includegraphics[width=\linewidth]{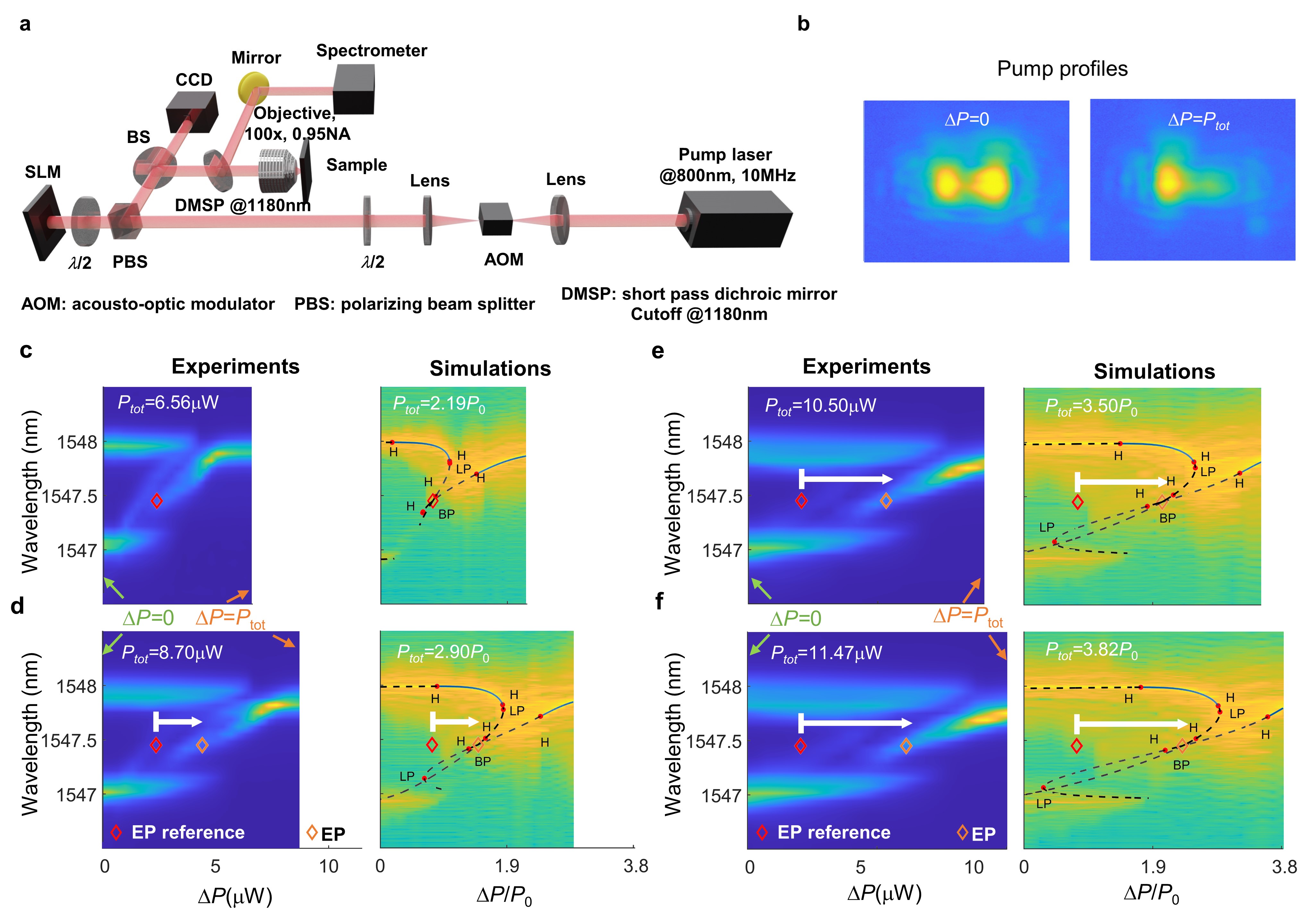}}
	\caption{\textbf{Tracking EPs above the lasing threshold.} (a) A schematic of the experimental setup used in this work, where an SLM is used to engineer the pump profile. (b) Examples of two pump patterns corresponding to the extreme cases of $\Delta P=0$ (both cavities are equally pumped), and $\Delta P=P_{tot}$ (only cavity 1 is pumped). (c)-(f) Experimental (left) and simulation (right panels) results for the lasing modes, characterized by their emission wavelength versus the pump imbalance $\Delta P$ for different total pump values $P_{tot}$. The diamonds represent the location of the EP as obtained from the expression $\Delta P=K(P_{tot}-2n_0\Gamma_{tot})/\kappa$ (see SM note \ref{Sec_DeltaP}). For reference, the EP from $P_{tot}=6.56\mu \text{W} $ is also indicated in all panels as a red diamond; the shift of the EP towards higher positive values of $\Delta P$ as $P_{tot}$ is increased can be clearly observed. Here the measured coupling and detuning are $K/\kappa=0.95$ and $\delta\omega/\kappa=-5.25$, respectively. For details on how the measurements were performed, see Methods. In the simulation, the detuning is chosen to be $\delta\omega/\kappa=2\alpha K=-5.70.$}
	\label{Fig_EP_Lasing}
\end{figure*}

\textit{Coupled photonic crystal nanolasers:---} In order to demonstrate the very different nature of EPs above the lasing threshold, we have fabricated a photonic crystal molecule similar to that shown in Fig. \ref{Fig_Schematic}, in which the cavity-to-cavity detuning is varied by means of the size of a side hole (yellow hole close to the left cavity). The pump profile in our experiment is controlled by using a spatial light modulator and the laser output is directed to a  spectrometer to measure the lasing frequency [Fig. \ref{Fig_EP_Lasing}(a) and (b)].

For typical values of $\alpha=2-5$, the condition $\delta \omega \sim -2K\alpha$ can be achieved with small coupling $K=0.13 \: \text{THz}$ and linear frequency shift of $\delta \omega=-0.74 \: \text{THz}$. The Q-factor for the dimer is $Q\approx 4200\, (1/\kappa\approx 7 \text{ps})$. To characterize the sample, we followed the procedure described in Ref. \cite{Hamel2015NP} (see also Methods). Note that, in this case, the yellow hole in Fig. \ref{Fig_Schematic}a is smaller that the background holes ($d=-0.1$), red-shifting cavity 1. The EP can be experimentally accessed above the lasing threshold provided: i) the total pump power exceeds twice the single laser threshold $P_0$ ($P_{tot}>2P_0$, see SM note S3); ii) introducing a positive pump imbalance $\Delta P$ between the two cavities that blue shifts cavity 1 with respect to cavity 2 so as to compensate for the external red-detuning. The non-Hermitian parameter $\Delta P$ can be continuously varied so as to approach $\Delta P|_{EP}$ (see detailed discussion in SM note \ref{Sec_DeltaP}). 

The left panels of Figs. \ref{Fig_EP_Lasing}(c)-(f) depict the experimental results characterizing the emission wavelength as a function of $\Delta P$ for different values of the total pump power. All the different cases share some generic qualitative behavior. Firstly, for $\Delta P=0$, there are two distinct spectral peaks indicating that the two modes of the photonic dimer are participating in the lasing action. 
This feature is consistent with the bifurcation diagrams (Figs. \ref{Fig_EP_Lasing} (c)-(f), right panels, where dashed lines account for unstable steady states), which predict that both modes are unstable for $\Delta P=0$ and give rise to mode beating limit cycles \cite{marconi2020mesoscopic} (Fig. S5(b), bottom panel, in the SM). The measured frequency splitting between the two modes is approximately $0.79 \: \text{THz}$. This compares well with the linear model, which predicts the presence of two supermodes oscillating at different frequencies with a splitting given by $\Delta \Omega=2\sqrt{(\delta \omega/2)^2+K^2}\sim 0.785 \: \text{THz}$. As $\Delta P$ is increased, multimode features appear in the laser spectra, which we relate to additional instabilities. Importantly, these are predicted by the model close to EP singularities, meaning that they can only observed under detuning compensation conditions, and therefore they can be taken as a signature of the proximity to the EP. Finally above a certain threshold for $\Delta P$, the  multimode emission collapses and only one lasing mode is observed.
The observed mode-structure is in good quantitative agreement with the numerical solutions of Eqs. (\ref{Eq-LaserRateEq}) with added noise terms [color maps in the right panels of Fig.\ref{Fig_EP_Lasing} (c)-(f)]. The details of how these solutions are obtained numerically are discussed in SM notes \ref{Sec_stability} and \ref{Sec_numerical_simulations}. Note that the theoretical plots indicate the presence of complex bifurcation structures that give rise to more than two steady-state solutions in certain parameter ranges. Another interesting generic observation from Fig. \ref{Fig_EP_Lasing} is that, as $\Delta P$ increases, the lasing emission in the PT-broken-like phase becomes red-shifted (e.g., Fig. \ref{Fig_EP_Lasing}(e), branch crossing the EP, from $\Delta P\sim 5 \mu$W to $10\mu$W). This is counter-intuitive since in this case, the applied pump to cavity one is increased. One thus may expect a blue shift due to the amplitude-phase coupling (i.e. the $\alpha$ parameter). A close inspection however, reveals that in this PT-broken-like phase, the lasing threshold is lower than the PT-unbroken-like phase. The gain clamping will thus results in smaller carrier concentration, and consequently a red shift from the operation in the PT-unbroken like phase as well as at the EP (see SM note \ref{Sec_EP_threshold} for a detailed discussion).\\

\begin{figure}[h]
	\centering
	{\includegraphics[width=\linewidth]{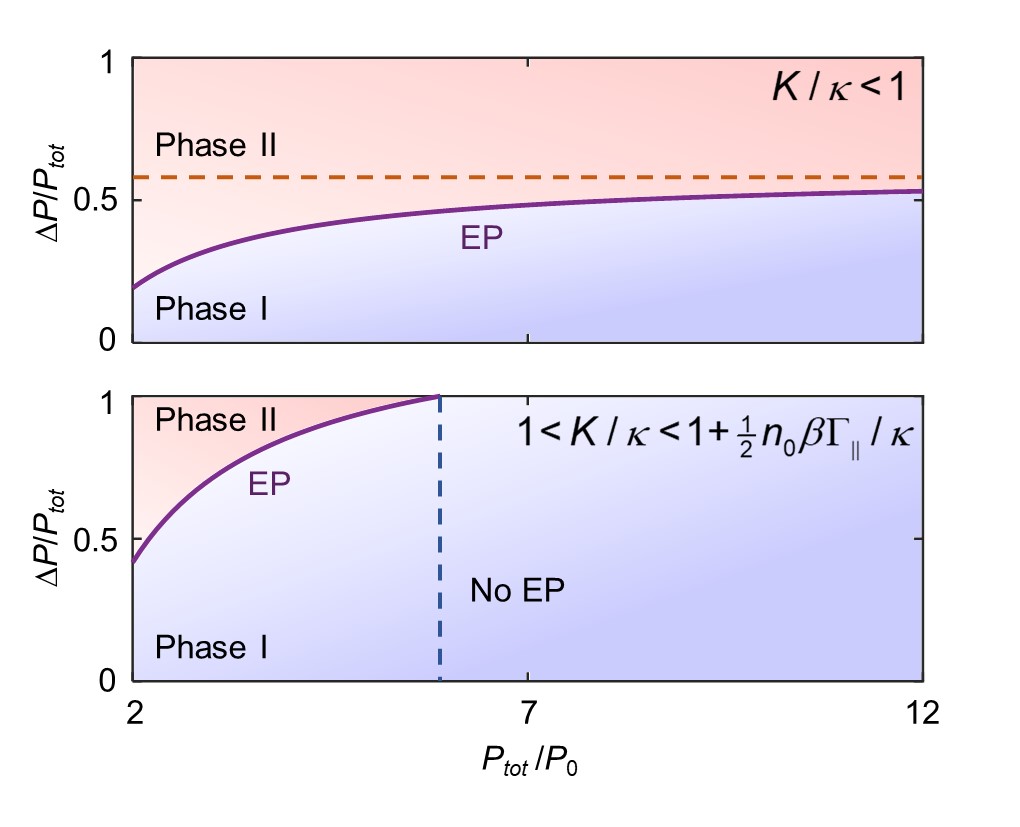}}
    \caption{\textbf{Different operation regimes} Top panel depicts the operating regimes when $K/\kappa<1$. In this case, exceptional points can be accessed above the lasing threshold for any value of $\Delta P_{EP}/P_{tot}$ but only for $\Delta P/P_{tot}<K/\kappa$ (the asymptotic red dashed line). On the other hand, as shown in the lower panel, when $1<K/\kappa<1+\frac{1}{2\kappa}n_0\beta\Gamma_{\parallel}$, the EPs can be accessed only in the domain defined by the condition $P_{tot}<\frac{2Kn_0\Gamma_{tot}}{K-\kappa}$ (see vertical dashed line). Here we denote Phase I as a PT-unbroken-like bimodal phase, and  Phase II as a PT-broken-like single mode phase (further details in the text). Finally, when $1+\frac{1}{2\kappa}n_0\beta\Gamma_{\parallel}<K/\kappa$, EPs cannot be accessed at all above the lasing threshold.}
	\label{Fig_PhaseDiag}
\end{figure}

In addition, contrary to what one would expect from using a simplified linear model, a close inspection of the experimental and theoretical data presented in Figs.\ref{Fig_EP_Lasing} (c)-(f)  reveals a rather interesting trend, namely that the location of this EP shifts to higher values of $\Delta P$ as the total pump power is increased. In this work, this feature presents a strong evidence supporting the presence of an EP above the lasing threshold. At the same time, this behavior also raises a question about how the regimes of operation vary as a function of total pump $P_{tot}$. Our analysis (see SM note S2) indicates that three distinct regimes can be identified based on the value of $K/\kappa$: (i) $K/\kappa<1$, (ii) $1<K/\kappa<1+\frac{1}{2\kappa}n_0\beta\Gamma_{\parallel}$, and (iii) $1+\frac{1}{2\kappa}n_0\beta\Gamma_{\parallel}<K/\kappa$. In the first of these regimes, the EP can be accessed above the lasing threshold for any value of $P_{tot}$ but below the horizontal line given by $\Delta P/P_{tot}<K/\kappa$ (see top panel of Fig. \ref{Fig_PhaseDiag}. In the second regime, the EP can be accessed only for a finite range for $P_{tot}$ as shown in the lower panel of Fig. \ref{Fig_PhaseDiag}. Finally, in the third regime, the EP cannot be accessed at all above the lasing threshold. Note that in Fig. \ref{Fig_PhaseDiag}, we denoted the areas below and above the exceptional line as phases I and II, respectively. In general, these phases cannot be associated with exact and broken PT phases, mainly due to the complex nature of the nonlinear bifurcation and the finite frequency shift due to the $\alpha$ factor. However, in cases where $\alpha= 0$ (such as the case in gas and rare-earth-doped solidstate lasers) or when the frequency shift due $\alpha$ is negligible compared to the coupling strength between the two resonators, one can make such a correspondence between phases I and II on one hand and the exact and broken PT phases on the other hand.

Next, we performed a second set of experiments to compare the behavior of the system when the EP is accessed above or below the lasing threshold. To this aim, here we fix the pumping rate of cavity one, $P_1$ and increase $P_2$. In the first case, we chose $P_1=1.1 P_0$. For this choice, the EP occurs at $P_{tot}<2P_0$. By recalling that the lasing threshold at the EP is exactly $2P_0$, it is clear that the EP in that case can be accessed only below the lasing threshold. As can be seen from the experimental results shown in the top panel of Fig. \ref{Fig_EP_Lasing}(a), under this condition, the system experiences laser self-termination and revival. Numerical calculations depicted in the lower panel also confirm these results. On the other hand, when the same experiment is repeated for $P_1=2P_0$, the EP is accessed above the lasing threshold and self-termination/revival behavior disappears, filling the laser extinction gap. These results are in good agreement with the linear model considered in Ref. \cite{Ganainy2014PRA} for analyzing laser self-termination. \\

Finally, we have also fabricated a second sample with a relatively weak frequency detuning of $\delta\omega/\kappa=1.19\ll 2K\alpha=22.38$ between the two photonic crystal cavities. This detuning cannot compensate for the carrier-induced frequency shift. Hence, asymmetric pump cannot be used to access an EP in this sample. As discussed before, in this case carrier-induced frequency shift will break the parity symmetry between the two cavities and as a result the system will not exhibit any EP. As a matter of fact, the experimental data in this case, which we present in Fig.~\ref{Fig_Laser_noEP} in SM note \ref{Sec_Extended}, reveal that the lasing characteristics are almost insensitive to $P_{tot}$. Importantly, the lack of compensation, hence the absence of a EP, does no impede transitions form two coexisting modes for small $|\Delta P|$, to a single localized mode for large $|\Delta P|$. While in many experimental examples such a mode transition is usually interpreted as a certain proximity to an EP, we stress the fact that such a phase transition, if any, instead of resulting from a weakly perturbed EP-bifurcation structure, takes place far from any EP, in the sense of Fig.~\ref{Fig_Bifurcation1}(c) [se also Figs. S7(b)]. 
\begin{figure}[t]
	\centering
	{\includegraphics[width=\linewidth]{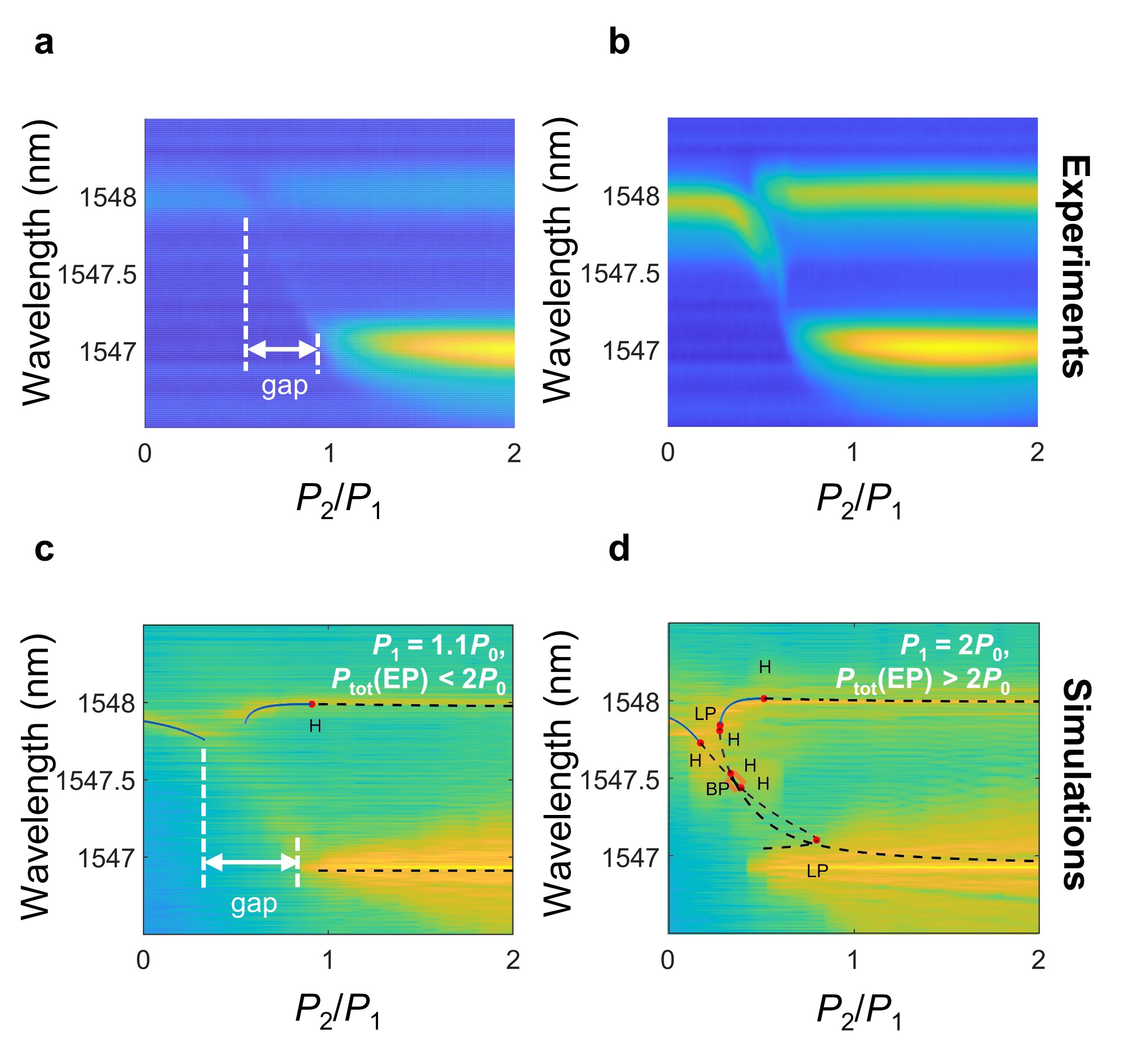}}
	\caption{\textbf{Lasing versus non-lasing EPs.} Experimental (top row) and theoretical (lower row) results for accessing EP below and above lasing threshold. In these figures, $P_1$ is kept constant and $P_2$ is varied. (a) $P_{tot}<2P_0$ ensures that the EP can be accessed only below the lasing threshold, as evidenced by the self-termination that takes place as $P_2/P_1$ is increased. (b) $P_{tot}>2P_0$, to ensure that the EP can be accessed above the lasing threshold. Here we observe mode switching without any self-termination effects. In all the figures, the color maps are presented in log scale for clarity.}
	\label{Fig_SelfTerm}
\end{figure}

\section*{Discussion}

While the recent interest in PT symmetry has attracted considerable attention, most of these studies have focused on utilizing EPs below the lasing threshold to engineer the cold cavity modes assuming that the character of the modes will remain intact above the lasing threshold, where the system is fundamentally linear. Even though this approach has proven useful in certain practical situations, it has two main drawbacks. Firstly, it ignores the rich dynamics that can arise due to the interplay between nonlinearity and non-Hermitian effects associated with EPs. Secondly, it may fail in systems where the addition of gain alter the nature of the modes (see for example Ref.~\cite{rivero2021non}). In the present work, we have bridged this gap by systematically demonstrating how EPs can be accessed and tracked above the lasing threshold in coupled semiconductor photonic crystal nanolasers. Contrary to previous studies that considered unavoidable cavity detuning as nuisance that impedes the formation of an EP above the lasing threshold~\cite{takata2021observing}, here we show that controllable detuning is actually a key ingredient for compensating the carrier-induced frequency shift and hence steering the system towards PT symmetry and EPs. Notably, our analysis shows that, in the nonlinear regime above the threshold, the EP becomes a nonlinear bifurcation point and solutions around it acquire a rich dynamical behavior. Namely, in some regions around this point, we find that all the steady state solutions are unstable and the lasing action becomes oscillatory. Such instabilities apply to the particular case a class-B laser systems ---to which semiconductor lasers belong---  (see Ref. \cite{https://doi.org/10.1002/lpor.202200377} and references therein), where the population inversion decay rate $\Gamma_{tot}$ is much smaller than the cavity damping rate $\kappa$. In turn, most laser models usually utilised in the literature so far for describing PT and EP-related phenomena are class-A laser systems, where the atomic population is assumed fast and it can therefore be adiabatically eliminated;  as a result, the EPs become stable steady states \cite{https://doi.org/10.48550/arxiv.2206.12969} which, as we show in this work, is a very different scenario from what is expected for (class B) semiconductor cavities.

A particularly interesting outcome of our work is the realization a certain constraint between the intercavity coupling and losses for the EP to form above the lasing threshold---a feature that does not have a counterpart in linear systems. While our work here focus in two single mode coupled cavities, it opens the door for future investigations on more complex laser systems. For instance, it is well known that interesting non-Hermitian effects can arise in deformed cavities having a large number of modes under passive conditions \cite{Wiersig2008PRA,Wiersig2011PRA}. Much less is known about the non-Hermitian effects in these systems in the nonlinear lasing regime. Similarly, understanding the interplay between non-Hermitian effects and disorder has only recently started to emerge~\cite{tzortzakakis2020non} but again in non-lasing setups. Extending this understanding to the nonlinear lasing regime can unlock more rich physics that so far has escaped attention. Finally we would like to comment on the lasing linewidth in our experiment. Even though we have not performed precise measurements of the emission linewidth, our experimental and numerical data in Fig.\ref{Fig_EP_Lasing} clearly demonstrate that the linewidth is finite in the presence of an EP. This in turn confirms the breakdown of the linewidth enhancement formula given by the Petermann factor \cite{petermann1979calculated,siegman1989excess,hamel1990observation,cheng1996experimental}, since the latter diverges at EPs. At the same time, it is clear that a relatively broader linewidth due to the overlap between different lasing modes can occur close the EP, which in fact is a nonlinear bifurcation point. This could make it more difficult to discriminate between the lasing frequencies which may degrade the operation of EP-based laser sensors. Furthermore, the instability of the steady state solutions at the EPs may pose an additional challenge for these sensors. In future works, we plan to investigate whether these features are generic in any coupled laser system operating at an EP or can be mitigated by tuning the design parameters and hence optimize the operation of these non-Hermitian sensors.

\section*{Methods}
Two coupled photonic-crystal cavities have been fabricated in an indium phosphide (InP) membrane (256nm) with four embedded InGa$_{0.17}$As$_{0.76}$P quantum wells.

The cavities are pumped using a pulsed laser ($\lambda=800$nm, 100ps duration and 10MHz repetition rate) to reduce thermal effect. The global intensity of the pump is controlled by an acousto-optic modulator (AOM). To control the pumps profile across the cavities in an independent manner, we use a spatial light modulator (SLM) to reshape the pump profile. The SLM is operated in amplitude modulation, in which we use two $\lambda/2$ plates to maximize both intensity and the contrast, respectively [The first $\lambda/2$ (close to the AOM) plate can maximize the intensity after the polarizing beam splitter; The second $\lambda/2$ (close to the SLM) rotates the polarization of the incident light to $45^\circ$ to achieve the higher contrast between the pump pattern and the unwanted background]. A infrared coated microscope objective with $\times$100 magnification and 0.95 numerical aperture is used to focused down the pump on the sample. The radiated emission is collected with the same objective and then spectrally resolved with a spectrometer.

The coupling and detuning are measured as follow,

i). Pump both cavity to obtain the split lasing modes $\omega_{\pm}$.

ii). Pump cavity one to obtain the blue-shifted frequency of cavity 1, $\omega_1$.

iii). Pump cavity two to obtain the blue-shifted frequency of cavity 2, $\omega_2$.

Together with the eigenvalues of the linear Hamiltonian, \begin{gather}
    \omega_+-\omega_-=\sqrt{4g^2+(\omega_1-\omega_2)^2}\\
    \omega_++\omega_-=\omega_1+\omega_2
\end{gather}

\section*{Acknowledgements}
L.G. acknowledges support by National Science Foundation under Grant No. PHY-1847240. R.E. acknowledges support from the Air Force Office of Scientific Research (FA9550-21-1-0202). K.J. acknowledges support from the China Scholarship Council (Grant No. 202006970015). 
This work is partially supported by the French National Research Agency (ANR), Grants No. ANR UNIQ DS078 and ANR-22-CE24-0012-01, the European Union in the form of Marie Sk\l odowska-Curie Action grant MSCA-841351, and by the RENATECH network.

\clearpage

\section*{Supplementary Material}
\renewcommand{\thetable}{S\arabic{table}}
\renewcommand{\thefigure}{S\arabic{figure}}
\renewcommand{\thesection}{S\arabic{section}}
\renewcommand{\theequation}{S\arabic{equation}}
\setcounter{equation}{0}
\setcounter{figure}{0}
\setcounter{section}{0}
\section{Compensating the carrier-induced frequency detuning}\label{Sec_Compensation}

As we mentioned in the main text, under asymmetric pumping of the two resonators,  the carrier-induced frequency shift (which is proportional to the linewidth enhancement factor $\alpha$) will break the parity symmetry between the two resonators. As a result, it is not possible to achieve PT symmetry and access the EP. In order to compensate for this dynamic effect, the two cavities must be designed to be initially asymmetric, i.e. having two different resonant frequencies. If the detuning is properly chosen, it can counterbalance the effect of the dynamic frequency shift. Here we derive the required frequency detuning to achieve this target. To do so, we start by considering the field equations of the laser rate model of Eq.(1a) in the main text, which can be expressed in the form: 

\begin{equation} \label{Eq_FieldEqs}
   \frac{d}{dt} \begin{bmatrix}
        a_1  \\
        a_2 
    \end{bmatrix}=
     \begin{bmatrix}
        i(\omega_1 + \Delta \omega_1) +g_1  & iK  \\
        iK  & i(\omega_2 + \Delta \omega_2) +g_2
    \end{bmatrix}
    \begin{bmatrix}
        a_1  \\
        a_2 
    \end{bmatrix}
\end{equation}

where $\Delta \omega_j=\frac{\alpha}{2}(n_j-n_0)\beta\Gamma_{\parallel}$  and $g_j=-\kappa+(n_j-n_0)\beta\Gamma_{\parallel}/2$, with $j=1,2$. Before we proceed, we emphasize the fact that the above equations incorporate the nonlinear effects arising from the coupling between the carrier and intensity. In other words, no linear approximations are made here. By expressing the fields as $a_j=A_j e^{i\phi_j} e^{i\Omega t}$ with $A_j$ and $\phi_j$ being real numbers, we arrive at: 

\begin{equation} \label{Eq_Eigvalues}
    \Omega_{\pm}=\omega_{avg} -i g_{avg} \pm \sqrt{K^2-\frac{(i\delta \omega_{12} +i\Delta \omega_{12}+\Delta g_{12})^2}{4}}
\end{equation}

where $\omega_{avg}\equiv (\omega_1+\omega_2+\Delta \omega_1 + \Delta \omega)/2$, $g_{avg}\equiv (g_1+g_2)/2$, $\Delta \omega_{12} \equiv \Delta \omega_1 - \Delta \omega_2=\frac{\alpha}{2}(n_1-n_2)\beta\Gamma_{\parallel}$, $\Delta g_{12} \equiv g_1-g_2$, and finally $\delta \omega_{12}\equiv \omega_1-\omega_2$. 

In order to steer the system to an EP, the following two conditions must be satisfied: (i) $\delta \omega_{12}=-\Delta \omega_{12}$, (ii) $\Delta g_{12}/2=K$. In writing this last relation, we assumed that $n_1>n_2$ since here $K>0$. Obviously, the other choice could have been made as well. Taken together, these two conditions lead to  

\begin{equation}
\label{Eq_EP_condition}
    \delta\omega_{12}= -2K\alpha.
\end{equation}

Importantly, since the lasing frequencies, $\Omega_{\pm}$ must be real, we find that at the EP (in fact even for $K>\Delta g_{12}$) $g_{avg}=0$, or equivalently $g_1=-g_2$. In other words, at the EP, the system respects PT symmetry.    

\begin{figure}[h]
	\centering
	{\includegraphics[width=\linewidth]{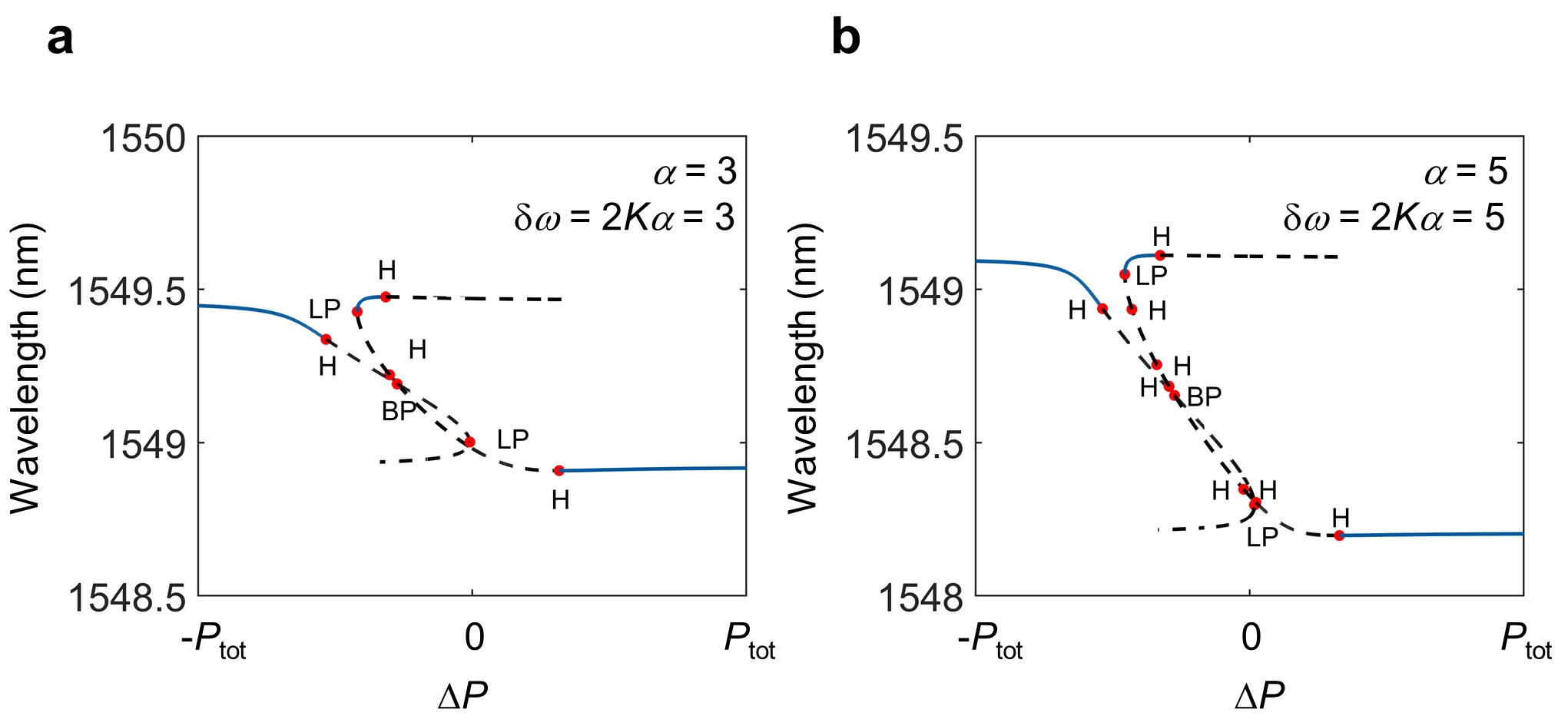}}
	\caption{Compensation of nonlinearity for different $\alpha$ values. $K=0.5, P_{tot}=3P_0$}
	\label{Fig_compensation}
\end{figure}

In order to confirm the validity of this analysis, we have calculated the lasing modes of two asymmetric laser cavities under the above derived detuning conditions. Figure.\ref{Fig_compensation} plots the resultant bifurcation diagrams for different values of $\alpha$ and the corresponding detuning. For both cases, effective PT symmetry is resorted at the branch points, which suggests that the compensation scheme described above can work under a wide range of parameters.

\section{\label{Sec_DeltaP}Locating the EP in the parameter space}
In this note, we derive the condition for operating at an EP above he lasing threshold. To do so, we start by recalling the characteristics of a lasing EP, which can be summarized as follows: (i) $\delta \omega=2K\alpha$, (ii) $g_1=-g_2$, (iii) The fields in both cavities have equal amplitude, and (iv) The phases of the fields in both cavities are different by a factor of $\pi/2$. As discussed before, the first of these conditions is the frequency detuning required to compensate for the carrier-induced frequency shift at the EP. When this condition is satisfied, the second relation becomes the lasing condition (imaginary part of $\Omega_{\pm}=0$) at the EP. The third and forth arise directly from the expression for the exceptional eigenvector associate with the solution of matrix eigenvalue problem in Eq.(\ref{Eq_FieldEqs}). In other words, we have: 

\begin{gather}
    \label{Eq_EP_net_gain}
    \frac{1}{2}(n_1-n_0)\beta\Gamma_{\parallel}-\kappa=-\left[\frac{1}{2}(n_2-n_0)\beta\Gamma_{\parallel}-\kappa\right]\\
    \label{Eq_EP_frequency}
    \frac{\alpha}{2}(n_1-n_0)\beta\Gamma_{\parallel}=\frac{\alpha}{2}(n_2-n_0)\beta\Gamma_{\parallel}+2K\alpha\\
    \label{Eq_EP_intensity}
    |a_1|^2=|a_2|^2,\\
    \label{Eq_EP_phase}
    \phi_1-\phi_2=\pi/2.
\end{gather}

By solving Eqs.~(\ref{Eq_EP_net_gain}) and (\ref{Eq_EP_frequency}), we can obtain the following expressions for the carrier numbers at EP:
\begin{gather}
    \label{Eq_EP_n1}
    n_1=n_0+\frac{2(K+\kappa)}{\beta\Gamma_{\parallel}},\\
    \label{Eq_EP_n2}
    n_2=n_0-\frac{2(K-\kappa)}{\beta\Gamma_{\parallel}}.
\end{gather}

Meanwhile, these carrier numbers can be related to the pump rates by solving the carrier rate equations (Eq.(1.b) in the main text) under steady states conditions, i.e., $\dot{n_1}=\dot{n_2}=0$. By doing so, we arrive at:

\begin{gather}
    \label{Eq_EP_pump_power}
    P_j-n_j\Gamma_{tot}=(n_j-n_0)\beta\Gamma_{\parallel}|a_j|^2,
\end{gather}

From Eq.~(\ref{Eq_EP_intensity}) and Eq.~(\ref{Eq_EP_pump_power}) we obtain:
\begin{equation}
    \label{Eq_EP_carrier_ratio}
    \frac{n_1-n_0}{n_2-n_0}=\frac{P_1-\Gamma_{tot}n_1}{P_2-\Gamma_{tot}n_2}.
\end{equation}

Finally, by combining Eqs.~(\ref{Eq_EP_n2}), (\ref{Eq_EP_pump_power}) and (\ref{Eq_EP_carrier_ratio}), we obtain:
\begin{gather} \label{Eq_DeltaP}
    \Delta P|_{EP}=\frac{K(P_{tot}-2n_0\Gamma_{tot})}{\kappa}
\end{gather}
where as in the main text, $\Delta P=P_1-P_2$. Note that the sign in front of the left hand side is a result of the assumption $n_1>n_2$ and the definition of $\Delta P$. Recalling the definition of the gain difference at EP $\Delta g_{2,1}|_{EP}=\beta\Gamma_{\parallel} (n_2-n_1)/2=2K$, we can rewrite Eq. (\ref{Eq_DeltaP}) as
\begin{equation} \label{Eq_DeltaP_DeltaG}
    \Delta P|_{EP}=-\frac{\Delta g_{2,1}|_{EP}(P_{tot}-2n_0\Gamma_{tot})}{2\kappa},
\end{equation}
which shows that the pump difference at the exceptional point, $\Delta P|_{EP}$, is not only a function of the gain difference, but also of the total pump power. At the laser threshold, it takes the simple form $\Delta P|^{th}_{EP}=(P_2-P_1)|^{th}_{EP}=2\Delta g_{2,1}\Gamma_{tot}/\beta\Gamma_{\parallel}$, which recovers the linear case.

On the other hand, to have EP in the strong pumping regime, Eq. (\ref{Eq_DeltaP}) must satisfy $\Delta P|_{EP}\leq P_{tot}$. In the high pumping limit it reduces to
\begin{equation}
\label{Eq_limit_DeltaP}
    \lim_{P_{tot}\rightarrow\infty}\frac{\Delta P_{EP}}{P_{tot}}=\frac{K}{\kappa},
\end{equation}
which implies that the system parameters must satisfy the condition $K<\kappa$, i.e. the coupling between the two cavities must be smaller than the loss factor of each cavity. Note that the condition $K<\kappa$ also arises from the less restrictive requirement of $n_{1,2}>n_0$, together with $K,\kappa>0$, as can be seen from Eqs. (\ref{Eq_EP_n1}) and (\ref{Eq_EP_n2}). \

\begin{figure}[t]
	\centering
	{\includegraphics[width=\linewidth]{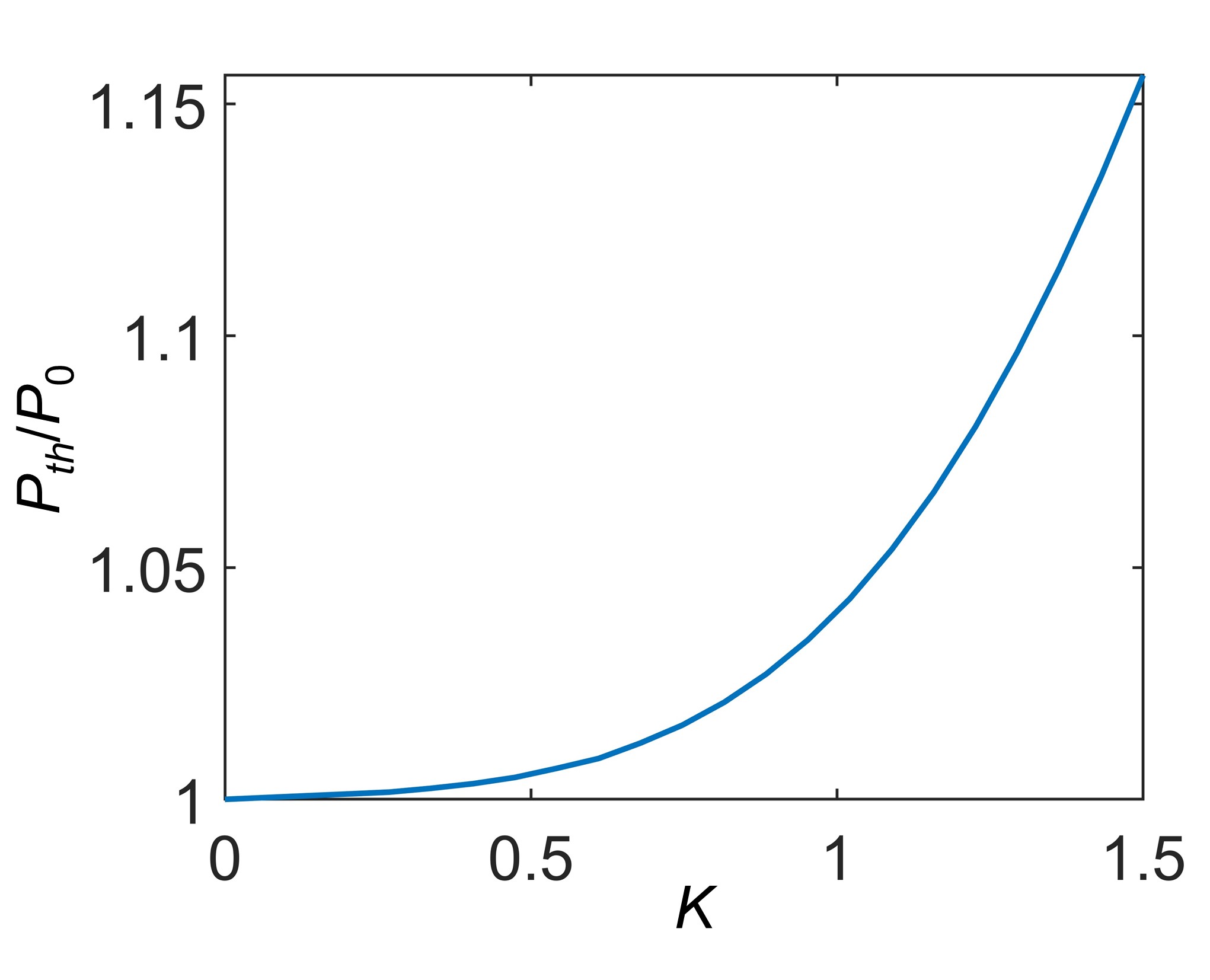}}
	\caption{Threshold of the broken PT phase.}
	\label{Fig_threshold}
\end{figure}

Finally, let us consider the extreme case where the EP is approached when one of the cavities is pumped,  namely $\Delta P|_{EP}=1$ with $P_1=P_{tot}=P_{EP}^{th}$, $P_2=0$ and $|a_1|^2=|a_2|^2=0$. In this case, the values of gain coefficients in both cavities are given by $g_1=-\kappa+\frac{1}{2}(n_1-n_0)\beta\Gamma_{\parallel}$ and $g_2=-\kappa-\frac{1}{2}n_0\beta\Gamma_{\parallel}$. The second of these formulas is a result of the fact that the steady state solution for the carrier in the second cavity  (see Eq.(1b) in the main text) under the conditions $P_2=0$ and $|a_2|^2=0$ is $n_2=0$. Note that the expression for $g_2$ indicates that cavity 2 is actually experiencing net loss. In fact this is the maximum possible loss, which partly due to optical loss and partly due to absorption in the QW layer. On the other hand, as we discussed before, the condition for lasing at an EP is $g_1=-g_2=K$. By combining these results together, it becomes obvious that the EP cannot be accessed above the laser threshold if $K>\kappa+\frac{1}{2}n_0\beta\Gamma_{\parallel}$. For the parameters used in out work (see SM note 6 for a full list of parameters), this becomes $K/\kappa=3.12$.

\begin{figure}[t]
	\centering
	{\includegraphics[width=\linewidth]{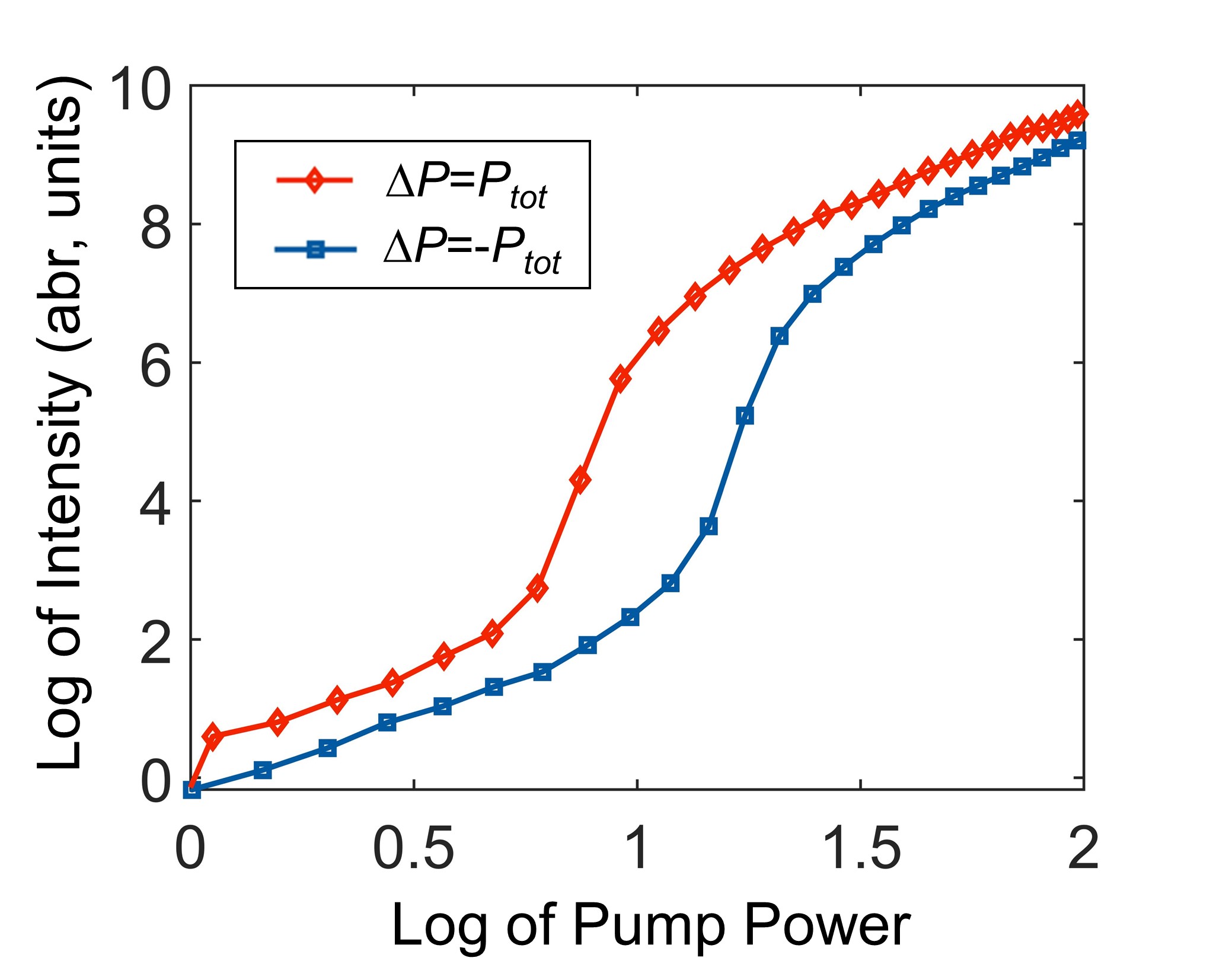}}
	\caption{Threshold of the of the system when a single cavity is pumped.}
	\label{Fig_logthreshold}
\end{figure}

\begin{figure*}[t]
	\centering
	{\includegraphics[width=\linewidth]{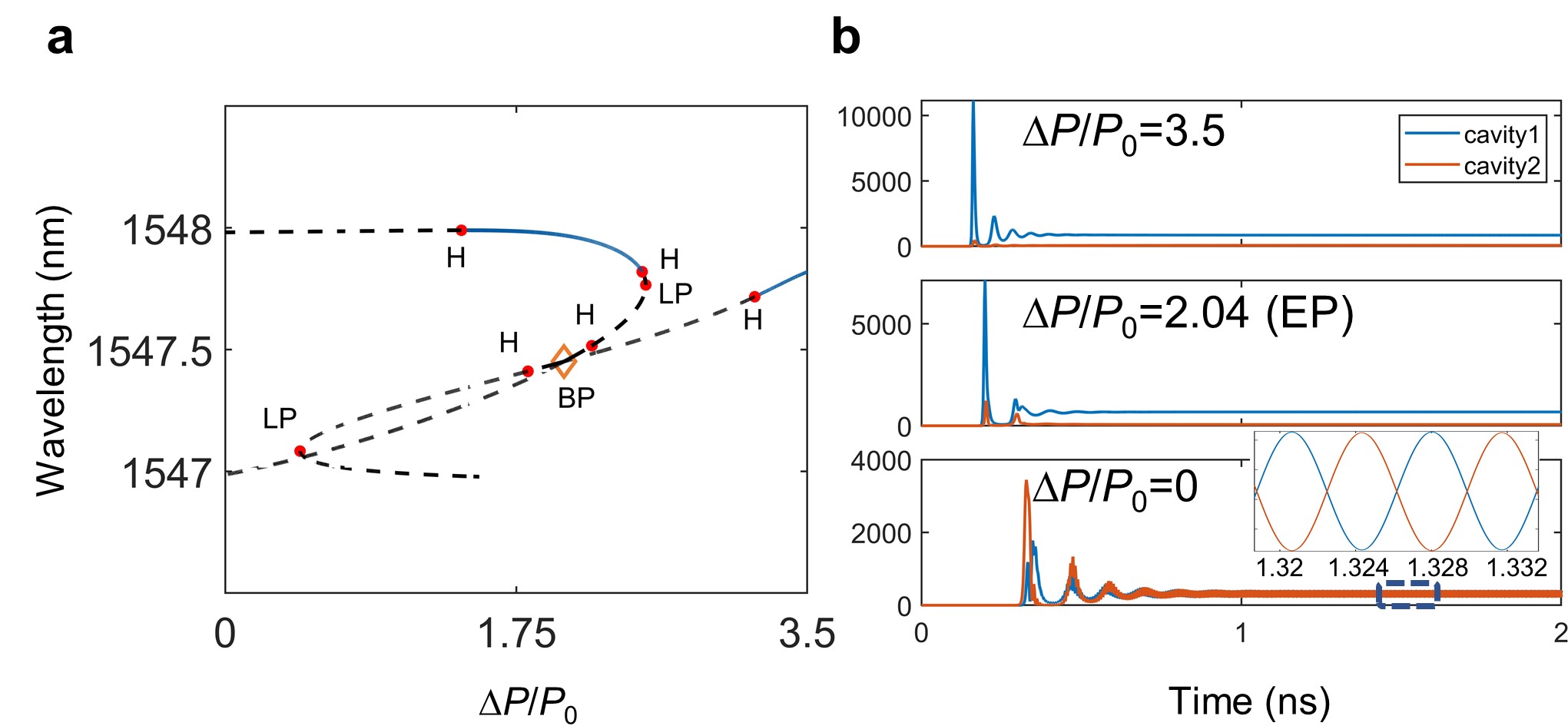}}
	\caption{\textbf{a} Bifurcation diagram and stability calculated for $P_{tot}=3.5P_0$. \textbf{b} Temporal dynamics for an exemplary stable solutions (top and middle panel) and one unstable, oscillator, solution (bottom panel).}
	\label{Fig_stability}
\end{figure*}

\section{\label{Sec_EP_threshold}Lasing threshold and frequency at EP}
At threshold, we have $|a_1|^2=|a_2|^2=0$. From Eq. (\ref{Eq_EP_pump_power}), we thus obtain:
\begin{gather}
    P_j^{th}=n_j\Gamma_{tot}.
\end{gather}
By using Eqs.~(\ref{Eq_EP_n1}) and (\ref{Eq_EP_n2}), find:
\begin{gather}
    P_{EP}^{th}=P_1^{th}+P_2^{th}=2(n_0+\frac{2\kappa}{\beta\Gamma_{\parallel}})\Gamma_{tot}=2P_0,
\end{gather}
where $P_0$ is the threshold of the single cavity laser. This last relation confirms that $P_{tot}>2n_0\Gamma_{tot}$ above the lasing threshold, which as expected results in a negative value for $\Delta P_{EP}$ (See Eq.(\ref{Eq_DeltaP})). To complete the discussion, we have also evaluated the lasing threshold in the broken PT phase when $\Delta P=-1$ numerically as a function of the coupling coefficient $K$. This result is shown in Fig.~\ref{Fig_threshold}.

To determine the lasing frequency at EP, we recall that at or above threshold $g_{avg}=0$,. Moreover, at EP, $\Omega_{\pm}=\omega_{avg}$. By recalling that $\omega_2=\omega_1+2K\alpha$ and that the lasing condition $g_1=-g_2$ implies that $\sum_{j=1}^{2} (n_j-n_0)\beta \Gamma_{\parallel}=2\kappa$, we find:

\begin{equation}
    \Omega_{EP}=\omega_1+\alpha(\kappa+K)
\end{equation}

\section{Light-in Light-out curve}
We experimentally measure the light-in vs light-out curve of our sample when two cavities are solely pumped (see Fig. \ref{Fig_logthreshold}). Note that the difference between two curves is caused by the alignment of the cavity, i.e., the spectrometer is aligned with respect to one of the cavities.

\section{\label{Sec_stability}Stability of the solutions}
To calculate lasing modes and their bifurcation diagrams, we express the complex fields as $a_j=A_j (t) e^{i\omega t+i\phi_j(t)}$. By substituting back in the laser rate equations (Eq.(1a) in the main text) we obtain:
\begin{gather}
    \label{Eq_amplitude1}
    \dot{A_1}=\frac{1}{2}\left[\beta\Gamma_{\parallel}(n_1-n_0)-2\kappa\right]A_1+KA_2\sin{\Delta\phi},\\
    \label{Eq_amplitude2}
    \dot{A_2}=\frac{1}{2}\left[\beta\Gamma_{\parallel}(n_2-n_0)-2\kappa\right]A_2-KA_1\sin{\Delta\phi},\\
    \begin{aligned}\label{Eq_phase}
         \dot{\Delta\phi}\equiv&\dot{\phi_1}-\dot{\phi_2}=\frac{1}{2}\alpha\beta\Gamma_{\parallel}(n_1-n_0)+\frac{A_2}{A_1}K\cos{\Delta\phi}-\\
         &\left[\frac{1}{2}\alpha\beta\Gamma_{\parallel}(n_2-n_0)+\frac{A_1}{A_2}K\cos{\Delta\phi}\right]+\delta\omega.
    \end{aligned}
\end{gather}

The lasing modes and their bifurcation diagrams are then calculated by solving these equations together with the carrier rate equations (Eq.(1b) in the main text) by using the open source continuation software package MatCont~\cite{dhooge2008new}. Accessing a particular mode using numerical integration of the laser equations can be achieved by choosing the proper initial noise in the system.

The stability of the modes are calculated by using linear stability analysis around the lasing modes, i.e. using the Jacobian matrix:
\begin{equation}
    \boldsymbol{J}=\begin{bmatrix}
        \frac{\partial f_{A_1}}{\partial A_{1}} & \frac{\partial f_{A_1}}{\partial A_{2}} &  \frac{\partial f_{A_1}}{\partial n_{1}} &
        \frac{\partial f_{A_1}}{\partial n_{2}} &
        \frac{\partial f_{A_1}}{\partial \Delta\phi} &\\
        \frac{\partial f_{A_2}}{\partial A_{1}} & \frac{\partial f_{A_2}}{\partial A_{2}} &  \frac{\partial f_{A_2}}{\partial n_{1}} &
        \frac{\partial f_{A_2}}{\partial n_{2}} &
        \frac{\partial f_{A_2}}{\partial \Delta\phi} &\\
        \frac{\partial f_{n_1}}{\partial A_{1}} & \frac{\partial f_{n_1}}{\partial A_{2}} &  \frac{\partial f_{n_1}}{\partial n_{1}} &
        \frac{\partial f_{n_1}}{\partial n_{2}} &
        \frac{\partial f_{n_1}}{\partial \Delta\phi} &\\
        \frac{\partial f_{n_2}}{\partial A_{1}} & \frac{\partial f_{n_2}}{\partial A_{2}} &  \frac{\partial f_{n_2}}{\partial n_{1}} &
        \frac{\partial f_{n_2}}{\partial n_{2}} &
        \frac{\partial f_{n_2}}{\partial \Delta\phi} &\\
        \frac{\partial f_{\Delta\phi}}{\partial A_{1}} & \frac{\partial f_{\Delta\phi}}{\partial A_{2}} &  \frac{\partial f_{\Delta\phi}}{\partial n_{1}} &
        \frac{\partial f_{\Delta\phi}}{\partial n_{2}} &
        \frac{\partial f_{\Delta\phi}}{\partial \Delta\phi} &\\
    \end{bmatrix}
\end{equation}
where $f_{\xi}$ defined as $\dot{\xi}=f_{\xi}(A_1,A_2,n_1,n_2,\Delta \phi)$ (for $\xi=A_j,n_j,\Delta \phi$), represent the fixed points of the rate equation, namely $\dot{\xi}=0,\dot{n_j}=0,\dot{\Delta\phi}=0$. Negative and positive values of the real part of the complex eigenvalues of the Jacobian then indicate stable and unstable solutions, respectively. 

Interestingly, the bifurcation diagrams in the main text and supplementary show that there exist a regime around the bifurcation point where all the modes are unstable. As illustrated in Figure.~(\ref{Fig_stability}), which is calculated for $P_{tot}=3.5P_0$, we found that in this regime, the modes do not admit steady state solutions by rather oscillate in time. 

\section{\label{Sec_numerical_simulations}Stochastic analysis of laser rate equations}
In the main text, we presented a comparison between the experimental results and numerical simulations obtained by solving the laser rate equations in the presence of stochastic noise. In general, there are different algorithms for integrating stochastic differential equations. Here we employed the Eurler-Maruyama method~\cite{kloeden1992stochastic}. The noise terms, which arise due spontaneous emission, was taken to be a white noise: $\langle F_{\mu}(t)F_{\nu}(t') \rangle=2D_{\mu\nu}\delta(t-t')$. The coefficient is $2D_{a_ia_i^*}=2D_{a_ia_i^*}=R_{sp}$, with $R_{sp}$ being the spontaneous emission rate $R_{sp}=\beta F_p B n_{1,2}^2/V_a$, where $F_p$ is Purcell factor, $B$ is the bimolecular radiative recombination rate and $V_a$ is the volume of the active medium~\cite{marconi2020mesoscopic}. The spectra are obtained after the Fourier transformation of the numerical integration. The numerical values of parameters used in the simulations are listed in table \ref{Table_parameters}.
\begin{table}[!h] 
\caption{Parameter values} 
\centering
\begin{tabular}{|p{1.in}<{\centering}|p{2.3in}<{\centering}|} 
  \hline
  \multicolumn{1}{|c|}{\bf Symbol} & \multicolumn{1}{|c|}{\bf{Values}}  \\ 
  \hline
  $\kappa$ & 140.86GHz \\ 
  \hline
  $\alpha$ & 3 \\ 
  \hline
  $\beta$ & 0.017 \\ 
  \hline
  $\Gamma_{\parallel}$ & 2.2GHz \\ 
  \hline
  $\Gamma_\text{tot}$ & 5GHz \\ 
  \hline
  $V_a$ & $0.016\times10^{-12}$cm$^3$ \\
  \hline
  $F_p$ & 1.03 \\
  \hline
  $B$ & 3$\times10^{10}$ cm$^3$s$^{-1}$\\
  \hline
\end{tabular}
\label{Table_parameters}
\end{table}

\section{\label{Sec_Extended}Extended data}
\begin{figure*}[t]
	\centering
	{\includegraphics[width=\linewidth]{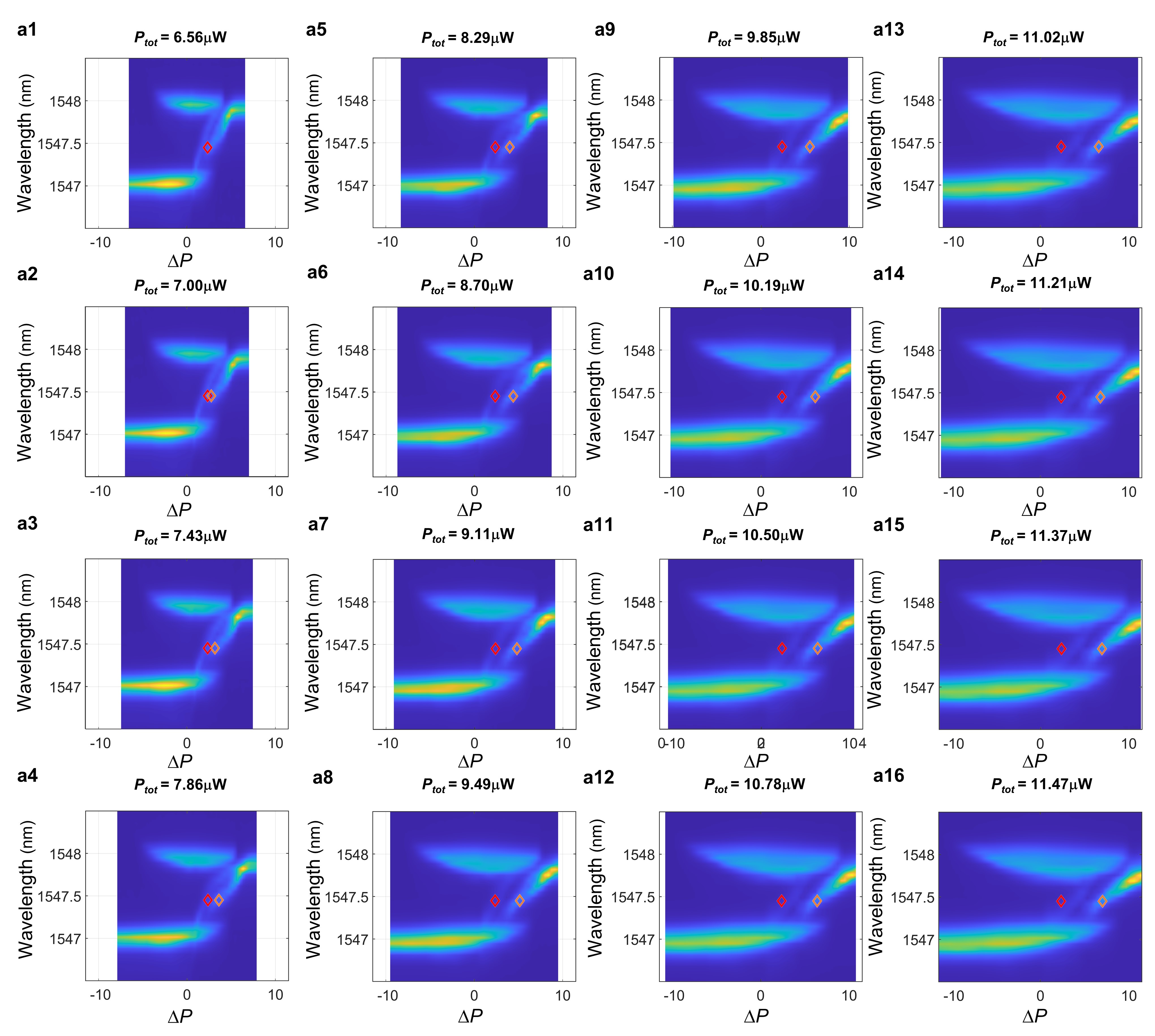}}
	\caption{Extended data for the measurements of the lasing modes as a function of $\Delta P$ for different values of $P_{tot}$. The same trend discussed in the main text is observed here as well, which supports the conclusion of this work.}
	\label{Fig_extended}
\end{figure*}

\begin{figure}[t]
	\centering
	{\includegraphics[width=3.4in]{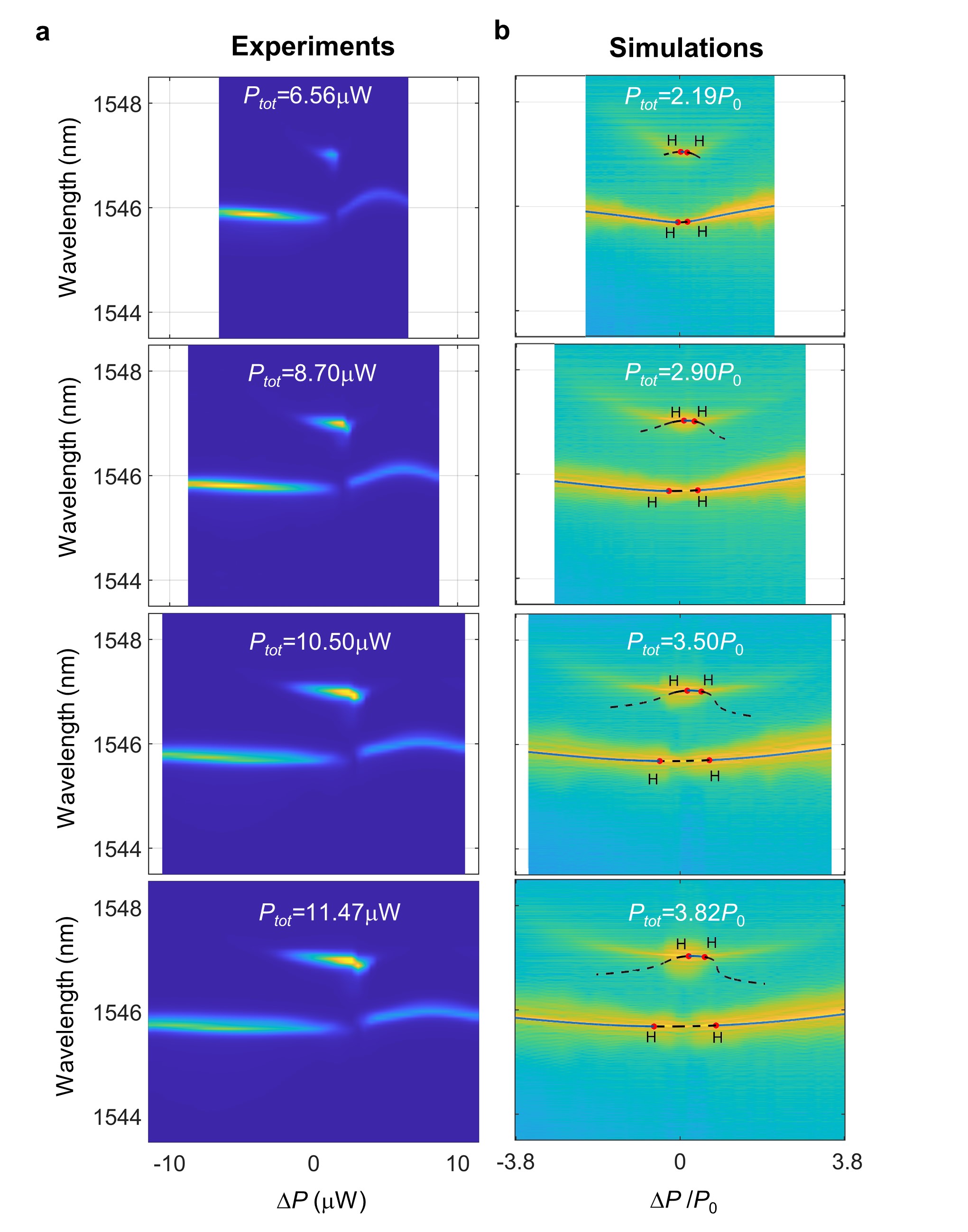}}
	\caption{Similar measurements to those presented in Fig.3 in the main text but for a different sample having $\delta\omega=1.19\ll 2g\alpha=22.38$ ($g=3.73$), i.e. with no frequency compensation to counterbalance the carrier-induced blueshift. As before, left panels present experimental data while right panels depicts theoretical results. In contrast to the behavior observed in Fig. 3 in the main text, here the lasing spectral pattern remains almost invariant as the total pump power is varied.}
	\label{Fig_Laser_noEP}
\end{figure}

For completeness, in this section, we present more measurement data for the lasing mode wavelengths as a function of $\Delta P \in [-1,1]$ evaluated at different values of $P_{tot}$. These results are plotted in Fig.\ref{Fig_extended}. By tracking the location of the bifurcation point in each figure, it is clear that the trend discussed in the main text (i.e. the shift of the bifurcation point toward larger negative values of $\Delta P$ as $P_{tot}$ increased does persist. On the other hand, Fig.\ref{Fig_Laser_noEP} depicts the experimental (top row) and numerical (lower row) results for the lasing characteristics of this system in the absence of frequency detuning compensation. In this case, there is no EP in the system and the lasing characteristics are considerably less sensitive to the total pump power.

\clearpage
\bibliographystyle{naturemag}

\end{document}